\documentclass[authoryear,preprint,12pt,3p]{elsarticle}
\usepackage[utf8]{inputenc}
\usepackage{t1enc}
\usepackage[dvipsnames]{xcolor}
\usepackage{float}
\usepackage{comment}
\usepackage{subcaption}

\usepackage{amssymb}

\usepackage[modulo]{lineno}

\biboptions{comma,round,sort}

\journal{Icarus}

\begin{document}

\begin{frontmatter}

\title{Deliquescence probability map of Mars and key limiting factors using GCM model calculations}

\author[label1,label2]{Bernadett D. P\'al\corref{cor1}}
\address[label1]{Research Centre for Astronomy and Earth Sciences, Konkoly Thege Miklos Astronomical Institute, ELKH, Budapest, Hungary}
\address[label2]{Eötvös Loránd University, Budapest, Hungary}

\ead{pal.bernadett@csfk.org}

\author[label1]{\'Akos Kereszturi}

\begin{abstract}
There may be a chance of small-scale ephemeral liquid water formation on present day Mars, even though the current climate does not support the existence of larger bodies of water. Through a process called deliquescence, hygroscopic salts can enter solution by absorbing water vapor directly from the atmosphere. Due to the absence of in-situ deliquescence experiments so far, the most reliable way to forecast deliquescence is through atmospherical modeling, however, the locations and times when salty liquid water could emerge are not yet well known. In this paper we present our results of likely brine formation on Mars, their proposed locations and seasons, as well as the possible limiting factors. For our calculations we used the data of Laboratoire de Météorologie Dynamique Mars General Circulation Model version 5. The results show that from L$_s$ 35$^\circ$ - L$_s$ 160$^\circ$, between 9 PM and 11 PM there is a good chance for calcium perchlorate deliquescence above 30$^\circ$ N, while in this zone the ideal regions are concentrated mostly to Acidalia Planitia and Utopia Planitia between 1 AM and 3 AM. We found that in the Southern Hemisphere, between L$_s$ 195$^\circ$ and L$_s$ 320$^\circ$, there is a noticeable, but weaker band in the vicinity of 60$^\circ$ S, and both Argyre Planitia and Hellas Planitia show some chance for brine formation. According to our statistics the key limiting factor of deliquescence could be relative humidity in most cases. Our results suggest that during summer -- early fall seasons, there could be deliquescence in both hemispheres in specific areas from the late evening until the early morning hours. There are only few studies detailing the geological and temporal distribution of brine formation through deliquescence, thus this work could be used as a good guide for future landing site analysis or in choosing a specific location for further research. \\

\end{abstract}

\begin{keyword}
deliquescence \sep Mars \sep near-surface relative humidity \sep general circulation model \sep liquid water
\end{keyword}

\end{frontmatter}


\section{Introduction}
\label{sec:intro}

Liquid water is one of the key elements in the search for extraterrestrial life. In the last decades Mars has been a primary target in finding signs of extant or ancient life. As we set our eyes on future crewed Mars explorations, the pursuit of liquid water is ever growing. However, the present martian conditions are not favourable for large scale pure liquid water formation on its surface \citep{haberle2001}; most of the studies favor the ephemeral emergence \citep[e.g.][]{brass1980, knauth2002, mellon2000, kereszturi2009, kossacki2004, mcewen2014, szynkiewicz2009}, but there are theoretical works suggesting a longer presence \citep[e.g.][]{clow1987, hecht2002}. The transient formation of salty brines is a likely candidate for water formation, as perchlorate salts have been identified by multiple martian rovers \citep{hecht2009det, navarro2010, leshin2013, ming2014}. Perchlorates are hygroscopic, meaning that they are able to form an aqueous solution if the temperature and relative humidity is appropriate through a process called deliquescence \citep{gough2011, gough2014, fischer2014, nuding2014, zorzano2009}. This process will be tested in-situ on board the ExoMars Rosalind Franklin rover (BOTTLE experiment \citep{torres2020}). One of the major topics investigated in this field is the habitability of brines, in which the most recent findings suggest a reason for optimism \citep{rivera2020, hallsworth2020}. \\

Perchlorates, including calcium perchlorate (Ca(ClO$_4$)$_2$) are abundant in the Martian top regolith \citep{hanley2012}, and might be a widespread component of the Martian soil, with 0.5-1\% concentration \citep{hecht2009}. This was identified by the Wet Chemistry Laboratory instrument onboard Phoenix mission \cite{toner2015}, and later detected by the SAM instrument onboard Curiosity rover \citep{navarro2013,glavin2013}. In the 1970's, it is proposed to have been detected by the Viking landers as well, and might be relevant to the field of astrobiology \citep{wadsworth2017}. Based on Earth based simulations and calculations, calcium perchlorate could contribute to daily hydration cycles and thus being influenced by the deliquescence process \citep{gough2019}. \\

Regarding the occurrence of calcium and magnesium perchlorate (Mg(ClO$_4$)$_2$) on Mars, the recent detections in the loose top regolith suggests a global occurrence, that is expected by wind mixing. Evaluating the formation conditions, calcium perchlorate might be produced by UV irradiance under Martian conditions with olivine minerals \citep{escamilla2020}, while atmospheric origin is also proposed \citep{catling2010} -- both models favour nearly global formation. If this salt have been produced by evaporation, it might show specific surface occurrence, however since there are no such observations the best current estimation is to consider a globally homogeneous presence with moderately small concentration (below 1\%). As a result our global spatial and temporal maps based distribution is a realistic approach using the present knowledge. \\

Our research focuses on the possibility of global-scale calcium perchlorate and magnesium perchlorate deliquescence. To our knowledge, there is only one study that attempted to study this phenomenon at this extent, published recently \citep{rivera2020}, using a different martian climate model. Therefore our proposing of another calculation method is beneficial to get a more complete picture of global deliquescence. The ultimate goal is to produce a map of possible thin liquid briny layer formation, that can be used as a guide in finding interesting locations, daily and seasonal periods for deliquescence studies, or in planning future Mars missions targeting liquid water. \\

To tackle this problem, we used an accomplished martian circulation model to simulate the annual variations of deliquescence likelihood on present Mars. We calculated the necessary environmental factors (relative humidity, temperature) for salt deliquescence and analyzed the global and annual characteristics. To demonstrate our results and find tendencies in brine emergence, we created zonally averaged and annual average global maps, as well as animations to show the day-to-day variations in detail (included as online supporting material). This paper summarizes our findings regarding magnesium perchlorate and calcium perchlorate deliquescence during the martian year 29. \\

\section{Methods}
\label{sec:methods}

We calculated the relative humidity using surface temperature, atmospheric pressure and water vapor volume mixing ratio data from the model Laboratoire de M\'et\'eorologie Dynamique Mars General Circulation Model (LMDZ GCM) version 5. This Mars climate model is detailed in \citet{hourdin2006}, the second generation of the model described in the works of \citet{sadourny1984} and \citet{forget1999}. The grid values were $64\times 48 \times 49$ in longitude, latitude, altitude respectively. For Mars, this $64\times 48$ horizontal grid corresponds to grid boxes of the order of $330\times220$ kilometers near the equator. The first layer of the altitude describes the first few meters above the ground. The data were simulated for the Mars year 29, which was a year without a global dust storm. For the simulation of this year, the model was set to the ,,climatology" dust scenario, which is built by averaging the krigged MY24-MY31 dust scenarios (excluding data from MY25 and MY28 global dust storms). For further details of the input files, please reach out to the authors. To derive the relative humidity with respect to ice and with respect to liquid, we calculated the saturation water vapor volume mixing ratio with an equation (used in the Martian Climate Database) based on the Goff-Gratch equation \citep{goff1946, list1984}. For the detailed calculations please refer to the methods section of \citet{pal2020}. \\

As the main local times of interest, we chose the late afternoon -- morning hours, from 5 PM to 7 AM, because during the day, with the rising surface temperatures, the relative humidity decreases to an almost zero level. Based on our earlier results \citep{pal2017} we chose to focus on the local times with a possible chance for deliquescence, and excluded those from this manuscript, that showed little to no chance in our selected time range. Finding ideal times is a difficult task, because relative humidity levels tend to increase during the night, when the temperatures are usually too low for deliquescence, and decrease during the day, when the temperatures are high enough \citet{pal2017}. In the late night and early morning hours, we see ideal time windows ranging from tens of minutes to a few hours at a time, thus these seem to be the best times to search for the emergence of liquid water. \\ 

We have analysed the chance for liquefaction of calcium perchlorate \citep{toner2014} and magnesium perchlorate \citep{mohlmann2011} by estimating the possibility of deliquescence on a binary scale (0 = no chance, 1 = definite chance). According to the research of \citet{rivera2018} if the relative humidity with respect to ice greatly exceeds 1, nucleation is favored rather than brine formation. Thus a certain time and place was determined ideal if both the temperature and the relative humidity with respect to liquid were above the necessary values, while at the same time the relative humidity with respect to ice stayed below 150\%. \\

\begin{table}[H]
    \centering
    \begin{tabular}{c c c} \\ \hline
       Salt  & Eutectic temperature & Water activity  \\ \hline
       Ca(ClO$_4$)$_2$ & 199 K & 0.51 \\ 
       Mg(ClO$_4$)$_2$ & 212 K & 0.53 \\ \hline
    \end{tabular}
    \caption{Parameters of the examined hygroscopic salts and water activity of the eutectic solution.}
    \label{tab:salts}
\end{table}

We calculated the relative humidity from GCM data and determined whether the necessary circumstances for deliquescence (see Table \ref{tab:salts}.) were present at a certain time and location. Since the model output is for 2 hour intervals, we calculated for every 2 hours between 5 PM -- 7 AM. We converted the results to NetCDF files, and used Panoply to visualize the data creating georeferenced maps at different local times and martian sols. For easier interpretation of our results we show zonal (Subsection \ref{subsec:zonal}) and annual averages (Subsection \ref{subsec:binary}), as well as animations in the online supporting materials. \\ 

\section{Results}
\label{sec:results}

The zonally averaged region with the greatest extent and highest average deliquescence probability, thus suggesting the best chance for deliquescence in the northern hemisphere, appears to be around 11 PM; while the best chance in the southern hemisphere is suggested to be around 1 AM. To determine the ideal time periods and locations, we used the GCM to model the Mars year 29 and calculated the probabilities on the binary scale described in Section \ref{sec:methods}. We show latitudinal zonal averages in Subsection \ref{subsec:zonal}. to highlight seasonal trends, while to illustrate the typical regions we show binary snapshots in Subsection \ref{subsec:binary}. To support the interpretation of the results, we investigate the annual climate of the polar regions in Subsection \ref{subsec:polar}, and highlight two of the most interesting seasons of the year, L$_s$ 90$$^\circ$$ and L$_s$ 270$$^\circ$$ (L$_s$ being the solar longitude) with global meteorological maps in Subsection \ref{subsec:globalclimate}. From the statistics of our results we focus on the likely limiting factors of deliquescence in \ref{subsec:limit}. Animations and figures excluded from this manuscript are included in the online version as supporting materials. \\

\subsection{Ideal seasons in a year - zonal averages}
\label{subsec:zonal}

In this subsection we show the most typical zonal average maps (average of given latitudinal bands) of deliquescence probability. During daytime the relative humidity is usually too low for deliquescence, resulting in scarce areas of possible liquid layer formation. The main point of the zonal averaged maps is to highlight how the martian seasons affect the chance for deliquescence, in different latitudes, throughout a year at a given local nighttime daily period. For the detailed, day-to-day latitude-longitude results we refer the reader to the animations in the online supporting materials. These longitudinally averaged maps highlight the seasons and latitudes, when there could be a chance for deliquescence, and do not depict specific geological locations. The notable moments are displayed on the maps: spring equinox (SE), summer solstice (SS), autumn equinox (AE) and winter solstice (WS), all with regards to the northern hemisphere.  \\


\begin{figure}[H]
    \centering
    \includegraphics[width=\linewidth]{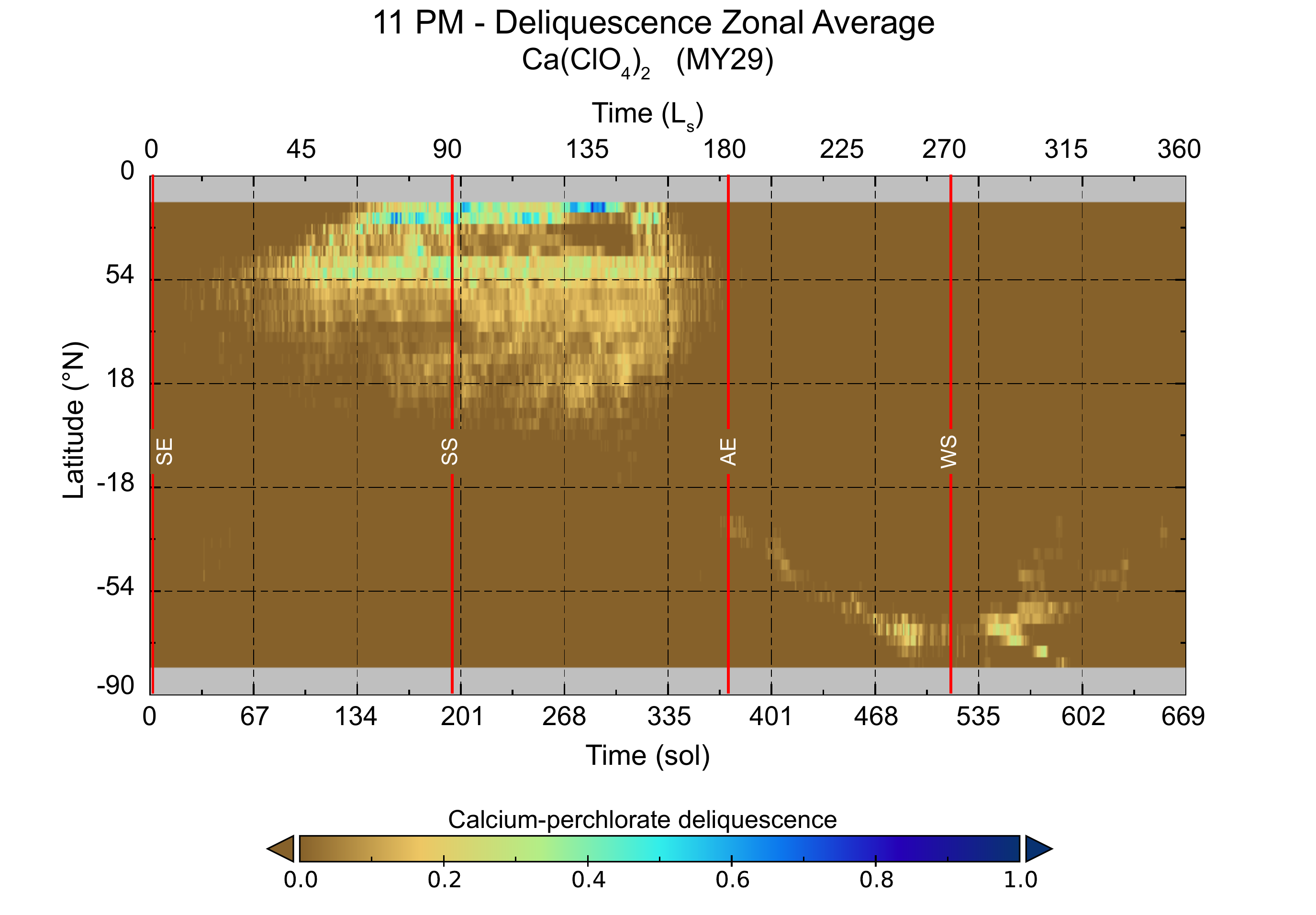}
    \caption{Zonal averages of calcium-perchlorate deliquescence probability at 11 PM local time during Mars year 29. Calculations are from LMDZ GCM, where 1 means definite chance and 0 means no chance.}
    \label{fig:11PMCa}
\end{figure}

The Northern mid and higher latitudes (20$^\circ$ and up) show a good chance of deliquescence during the northern late spring and summer to early autumn (L$_s$ $\approx$ 80 - 140$$^\circ$$). In Figure \ref{fig:11PMCa}. we can see the zonal averages of binary deliquescence possibilities during the Martian year 29 at 11 PM local time. The prominent area in the northern hemisphere is the ``teardrop" shaped zone between approximately sol 40 - 360. The saturation above 75$^\circ$N is most likely influenced by the perennial H$_2$O icecap layer. The upper arch of the northern area starting from sol 67 is in agreement with the seasonal sublimation of the CO$_2$ ice cap, exposing the H$_2$O ice underneath during the retreat of seasonal CO$_2$ ice cap. The higher temperatures and higher local atmospheric water vapor lead to a good chance for deliquescence, according to the model. As the year progresses, from sol 200, we see a steady expansion towards the southern hemisphere. The thin ``line" remains the sole feature from sol 360 until the end of the year, steadily shifting towards south until sol 600. At the end of the year, the candidate latitudes for deliquescence seem to shift back towards the equator. \\ 


\begin{figure}[H]
    \centering
    \includegraphics[width=\linewidth]{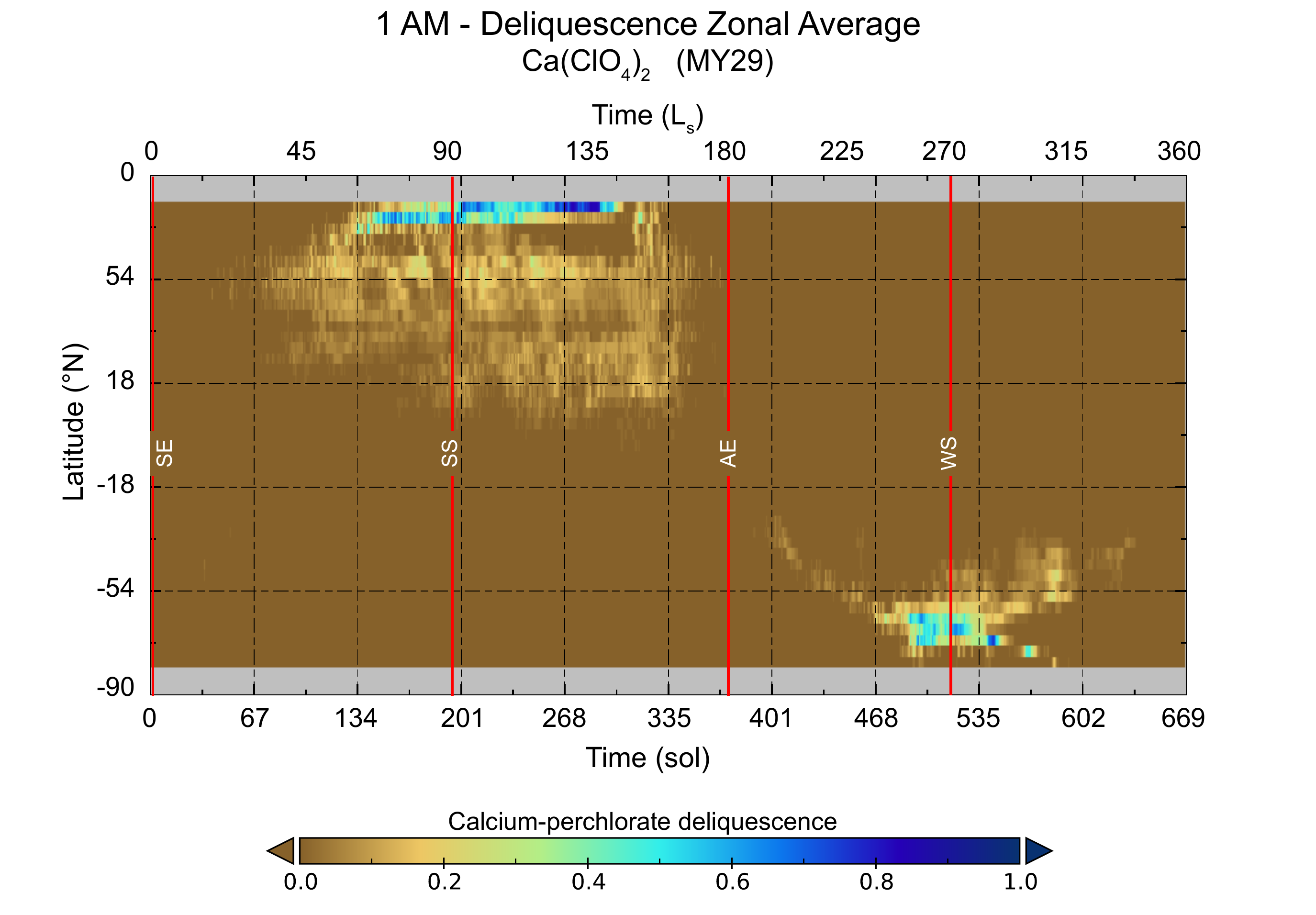}
    \caption{Zonal averages of calcium-perchlorate deliquescence probability at 1 AM local time during Mars year 29. Calculations are from LMDZ GCM, where 1 means definite chance and 0 means no chance.}
    \label{fig:1AMCa}
\end{figure}

The chance for deliquescence in the southern hemisphere peaks at 1 AM (Figure \ref{fig:1AMCa}., compared to all the other local times shown in the online supporting materials), though the areal extent is still significantly smaller than in the northern counterpart. There is a rather thin line starting from around sol 400, branching out into an intrigue shape as the year progresses. Just as in the northern case, the edge of the potentially ideal region is in good agreement with the receding polar CO$_2$ ice cap. As the CO$_2$ ice sublimes, the now exposed H$_2$O ice heats up and sublimes quickly, opening a very brief window of warm temperatures and higher local water vapor levels. \\




\begin{figure}[H]
    \centering
    \includegraphics[width=\linewidth]{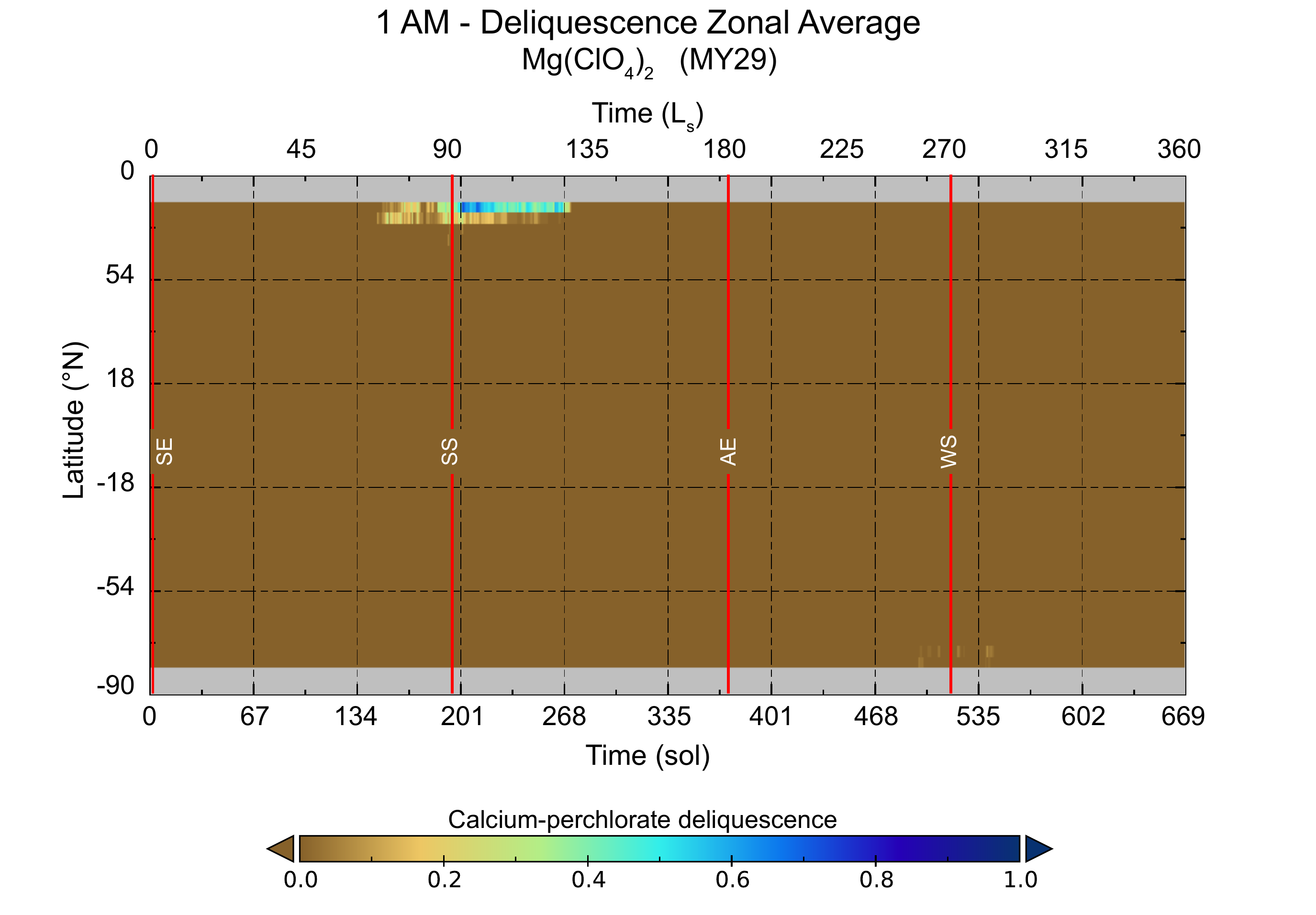}
    \caption{Zonal averages of magnesium-perchlorate deliquescence probability at 1 AM local time during Mars year 29. Calculations are from LMDZ GCM, where 1 means definite chance and 0 means no chance.}
    \label{fig:1AMMg}
\end{figure}

The magnesium perchlorate shows little chance for deliquescence throughout the Martian year 29 compared with calcium perchlorate. The visible area in Figure \ref{fig:1AMMg} is limited between sol 135 - 270 and is confined to the polar regions, where the perennial water ice may be influencing the calculations. There is a barely visible scatter between sol 490 - 540 near the southern polar region, however the rest of the latitudes remain empty. This 1 AM data is the best chance we have seen in our magnesium perchlorate results. \\

\begin{figure}[H]
    \centering
    \includegraphics[width=\linewidth]{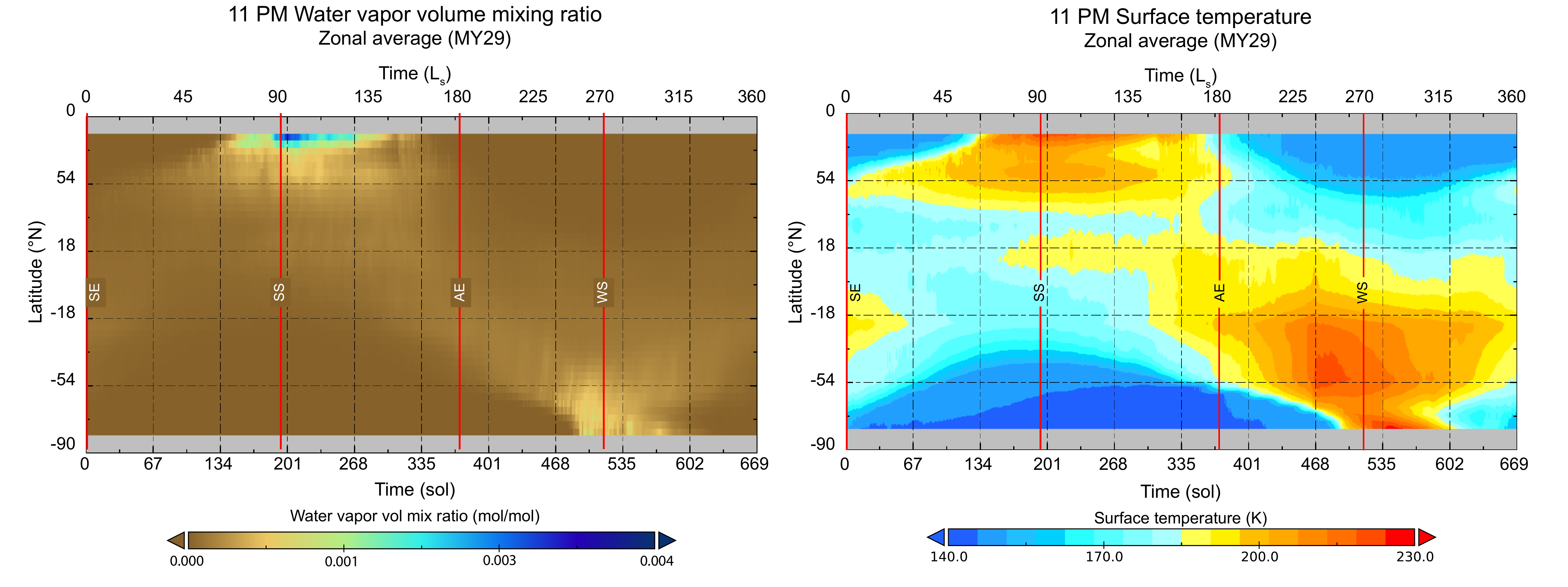}
    \caption{Zonal averages of water vapor volume mixing ratio and surface temperature at 11 PM local time during Mars year 29.}
    \label{fig:11PMtw}
\end{figure}

The annual "wave-like" tendency seen in the zonally averaged deliquescence maps (Figure \ref{fig:11PMCa}, \ref{fig:1AMCa}.) is clearly observable in the atmospheric water vapor and surface temperature figures (Figure \ref{fig:11PMtw}) too. The regions with higher water vapor and higher temperatures are the ones that show up potentially ideal in the deliquescence maps, as expected. \\

\subsection{Annual climate at the Polar regions}
\label{subsec:polar}

We have seen in the previous results, that the edges of the zonally averaged, potentially ideal regions follow the receding CO$_2$ ice caps in both hemispheres. However, while in the north, the likely ideal periods span right up to the poles, in the south, the shape stops around 75$^\circ$S. To better understand this, we examine the annual climate of the higher polar regions below. \\

\begin{figure}[H]
    \centering
    \includegraphics[width=\linewidth]{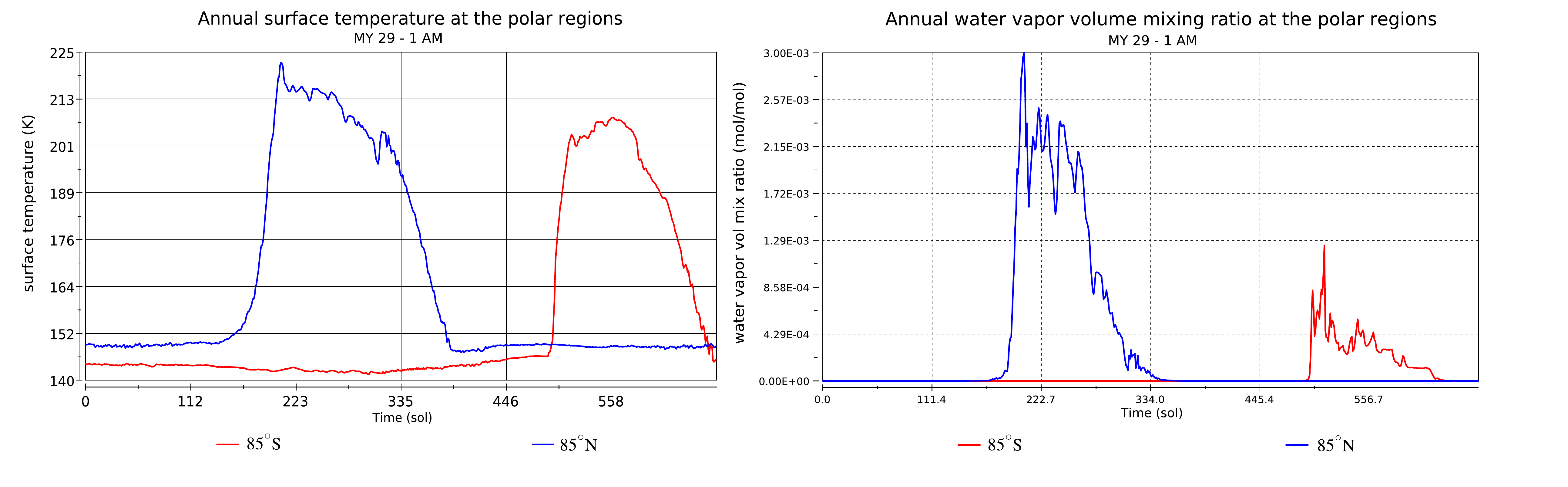}
    \caption{Annual variation of the surface temperatures (left) and the water vapor volume mixing ratio (right) at the northern (blue) and southern (red) polar regions during the martian year 29.}
    \label{fig:tswvappoles}
\end{figure}

First we examine the annual temperature and water vapor variability, to figure out, why the northern polar region shows a chance for deliquescence, while the southern does not. In Figure \ref{fig:tswvappoles}. we see the annual surface temperature curves at 85$^\circ$ S and 85$^\circ$ N with red and blue colors respectively. The length of the northern and southern summers do not differ as much as to explain the difference in deliquescence. In the northern polar region the maximum temperatures reach up to 223 K, while these remain lower, around 210 K in the southern hemisphere; however, both go above the minimum eutectic temperature of calcium perchlorate. \\

The average annual water vapor volume mixing ratio maximum is lower in the southern hemisphere, reaching only about 33\% of the average northern maximum. From this graph it seems, that during the local summer, the northern polar cap evaporates stronger, thus the amount of water vapor in the near-surface atmosphere is also higher. This combined with the high enough temperatures could lead to better chances for deliquescence in the northern polar region. \\

Another main difference between the northern and southern seasonal ice caps is their composition. The measurements of the OMEGA imaging spectrometer confirmed that during the recession state, the southern seasonal polar ice cap is mainly composed of CO$_2$ ice  \citep{langevin2007}. There are only a few specific locations, where surface H$_2$O ice is present between around 220$$^\circ$$ and 250$$^\circ$$ L$_s$. In agreement with this, the occasional deliquescence chance close to the southern pole is visible in the animations, but not enough to show up in the averaged maps. The seasonal northern polar cap however contains both CO$_2$ and H$_2$O ices \citep{schmitt2006}. During the local summer and the recession of the polar caps it is thus expected, that the northern polar regions would show a higher chance of deliquescence. \\

\subsection{Ideal geological regions - annual probability}
\label{subsec:binary}

In this subsection we show global annual averages of the deliquescence probabilities for Mars year 29 to highlight the potentially interesting geological regions. The maps shown in this section are not averaged longitudinally anymore, instead we averaged the results of the whole martian year 29. Please note, that the average chance for deliquescence ranges from 0 to 0.3 instead of 0 to 1 for better visibility. For the detailed day-to-day data before averaging, as well as for other local times, we refer the reader to supplementary animations and figures in the online version. \\


\begin{figure}[H]
    \centering
    \includegraphics[width=\linewidth]{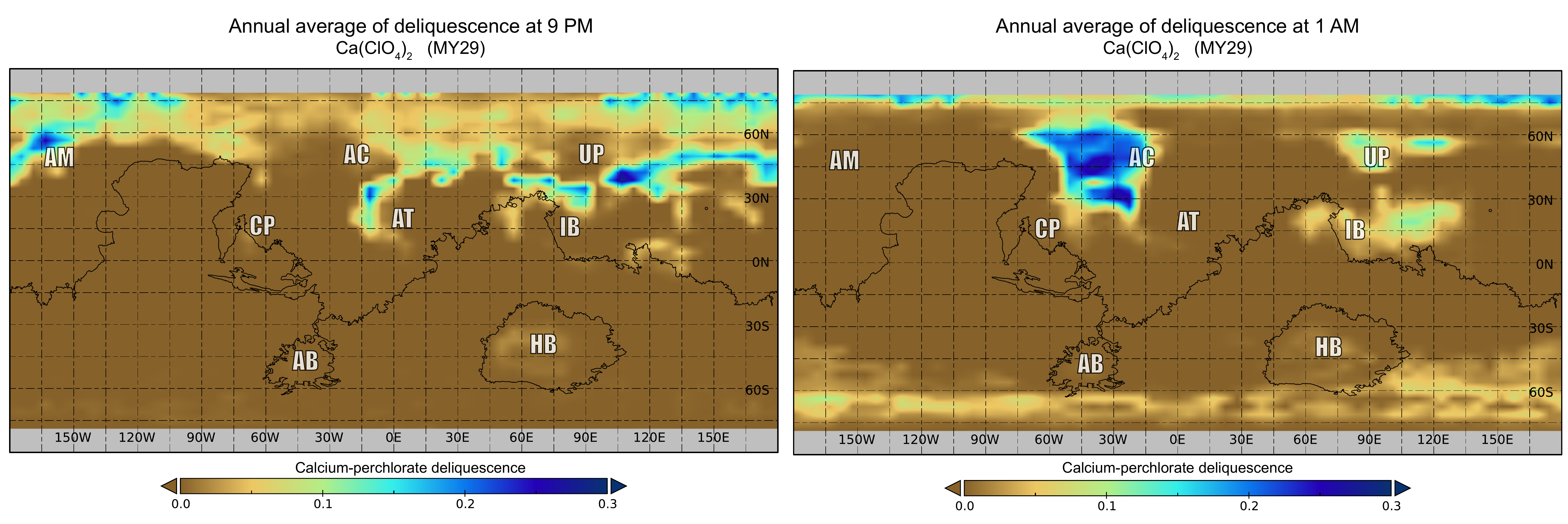}
    \caption{Annual averages of calcium-perchlorate deliquescence at 9 PM (left) and 1 AM (right). Data is from LMDZ GCM for Mars year 29. On the scale 0 means no chance and 1 would mean definite chance.}
    \label{fig:Ca1-3AMavg}
\end{figure}

 At 9 PM the latitudes above 30$^\circ$ N present the overall best chance for deliquescence throughout the Martian year 29, as seen in Figure \ref{fig:Ca1-3AMavg}. (left). Here we chose to include the 9 PM local time map instead of the 11 PM (online supporting material), because the contrast between 9 PM and 1 AM is more apparent. In Acidalia Planitia, around 15$^\circ$ W there is a region with elevated chance of deliquescence, which stretches down nearly to the equator. Utopia Planitia also strikes out as a good place for deliquescence at this time. We can see fainter regions southeast of Isidis Basin, at the equator between 105 - 135$^\circ$ E and in Hellas Planitia. Around 1 AM the northern upper latitudes show much concentrated possible deliquescence regions in Acidalia Planitia and Utopia Planitia. The somewhat continuous band between 45$^\circ$ - 75$^\circ$ N is barely visible, however, the southern band around 60$^\circ$ S is brighter. Hellas Planitia and Argyre Planitia both show a slight chance of deliquescence (this is a bit hard to see, we advise the reader to see the online supporting animations).  \\ 
 
 \begin{figure}[H]
    \centering
    \includegraphics[width=\linewidth]{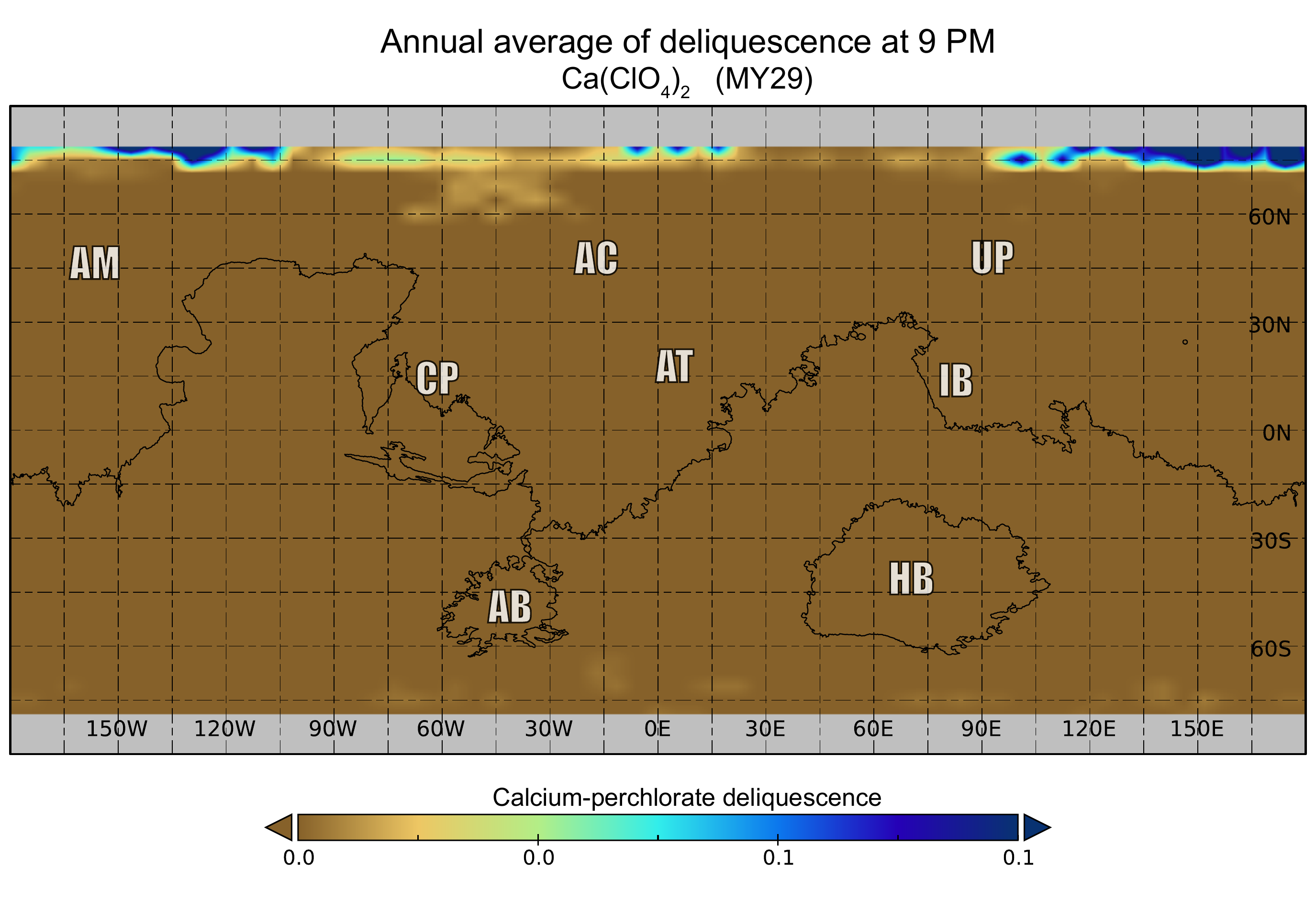}
    \caption{Annual average of magnesium perchlorate deliquescence at 11 PM. Data is from LMDZ GCM for Mars year 29. On the scale 0 means no chance and 1 would mean definite chance. Please note, that the scale is different from the rest to better visualize the small chance of deliquescence.}
    \label{fig:Mg11PMavg}
\end{figure}

There is very little chance for magnesium-perchlorate deliquescence according to our calculations. If we look at Figure \ref{fig:Mg11PMavg}. we can see, that there are barely any regions visible outside of the vicinity of the northern polar cap, above 75$^\circ$ N. Please note, that unlike the previous figures, the scale here ranges only from 0 to 0.1 in order to make the area between 75$^\circ$ W to 30$^\circ$ W, 60$^\circ$ N - 75$^\circ$ N visible. That zone in the northern parts of Acidalia Planitia show the only possible chance for deliquescence throughout the year. We chose 11 PM to show as the annual average, because this showed the overall best chance in our results. There are some spots below 60$^\circ$ S, but they form no continuous band, nor do they show a significant possibility. \\
 
 \subsection{Global climate modelling results}
 \label{subsec:globalclimate}
 
 Looking at the zonally averaged maps L$_s$ 90$^\circ$ and L$_s$ 270$^\circ$, 1 AM local time seem to be two of the best times for deliquescence in the northern and southern hemispheres respectively. In Section \ref{subsec:binary}. we have shown the annual averaged maps, detailing the global regions with the best chance for deliquescence. In this subsection we included global meteorological maps for L$_s$ 90$^\circ$ and L$_s$ 270$^\circ$, to aid the interpretation of our deliquescence probability results. \\

\begin{figure}[H]
    \centering
    \includegraphics[width=\linewidth]{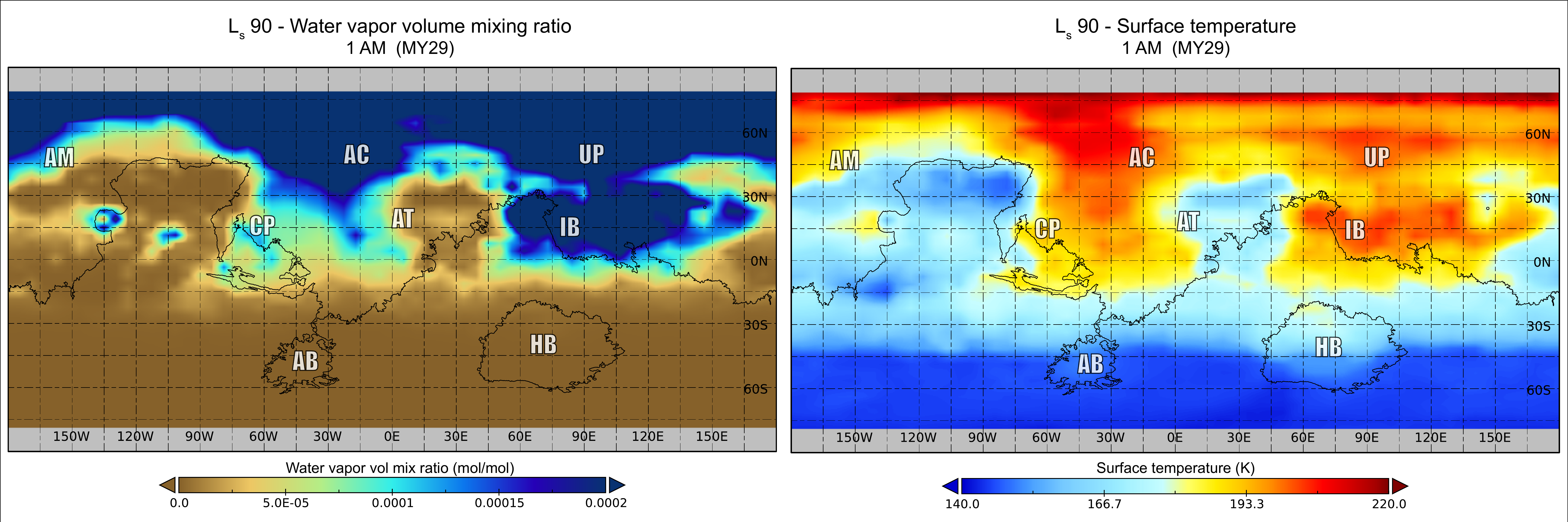}
    \caption{The global water vapor volume mixing ratio (left) and surface temperature (right) at L$_s$ 90$^\circ$ ($\approx$ sol 195) 1 AM local time for Mars year 29. }
    \label{fig:tswvapls90}
\end{figure}

With regard to seasons, in Figure \ref{fig:1AMCa}. L$_s$ 90$^\circ$ ($\approx$ sol 195) is an interesting point with the deliquescence chance reaching down almost to the equator, while the polar regions are not yet saturated. With regard to location, in the 1 AM annual average figures (Fig. \ref{fig:Ca1-3AMavg}., right) Acidalia Planitia, Utopia Planitia and the region around Isidis Basin strike out as the regions with the highest average deliquescence chance. At L$_s$ 90$^\circ$ both the water vapor ratio and the surface temperature is high enough at these regions (Fig. \ref{fig:tswvapls90}.). In the left figure we can see, that the water vapor content is similarly high around 60$^\circ$ N, however if we take the surface temperature (right side) into consideration, the regions with higher deliquescence chance have approximately 15-20 K higher temperatures at 1 AM. \\

\begin{figure}[H]
    \centering
    \includegraphics[width=\linewidth]{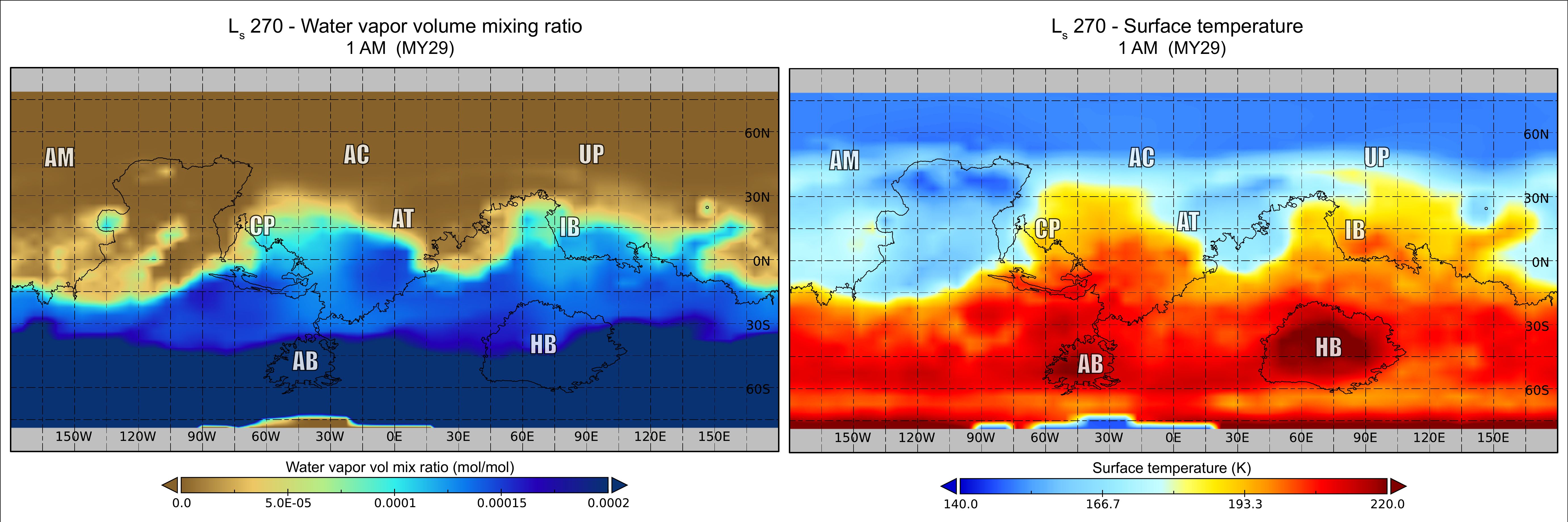}
    \caption{The global water vapor volume mixing ratio (left) and surface temperature (right) at L$_s$ 270$^\circ$ ($\approx$ sol 515) 1 AM local time for Mars year 29.}
    \label{fig:tswvapls270}
\end{figure}

The case of L$_s$ 270$^\circ$ ($\approx$ sol 515) is similar to L$_s$ 90$^\circ$. In the zonal average maps (Fig. \ref{fig:1AMCa}.) this season shows the best chance for deliquescence in the southern hemisphere. In the annual average maps (Fig. \ref{fig:Ca1-3AMavg}.) the southern hemisphere seems to show a significantly lower chance overall, than the northern hemisphere. Still, three regions strike out as possibly ideal locations: Argyre Basin, Hellas Basin, and the somewhat continuous band around 60$^\circ$ S. In the L$_s$ 270$^\circ$ snapshots, shown in Figure \ref{fig:tswvapls270}., we can see that almost the entire southern hemisphere shows elevated water vapor levels. Please note, that the water vapor range is smaller, than in the northern hemisphere to be able to visualize finer details. The atmosphere above Argyre Basin shows higher water vapor content, than the northern parts of Hellas Basin. Both of these regions have elevated temperatures, compared with their surroundings. While the water vapor content is similar, the temperatures differ in the 60$^\circ$ S band; but still seem to stay just above the eutectic temperature threshold of calcium perchlorate. \\


\subsection{Main limiting factors of deliquescence}
\label{subsec:limit}

\begin{figure}
    \centering
    \includegraphics[width=\linewidth]{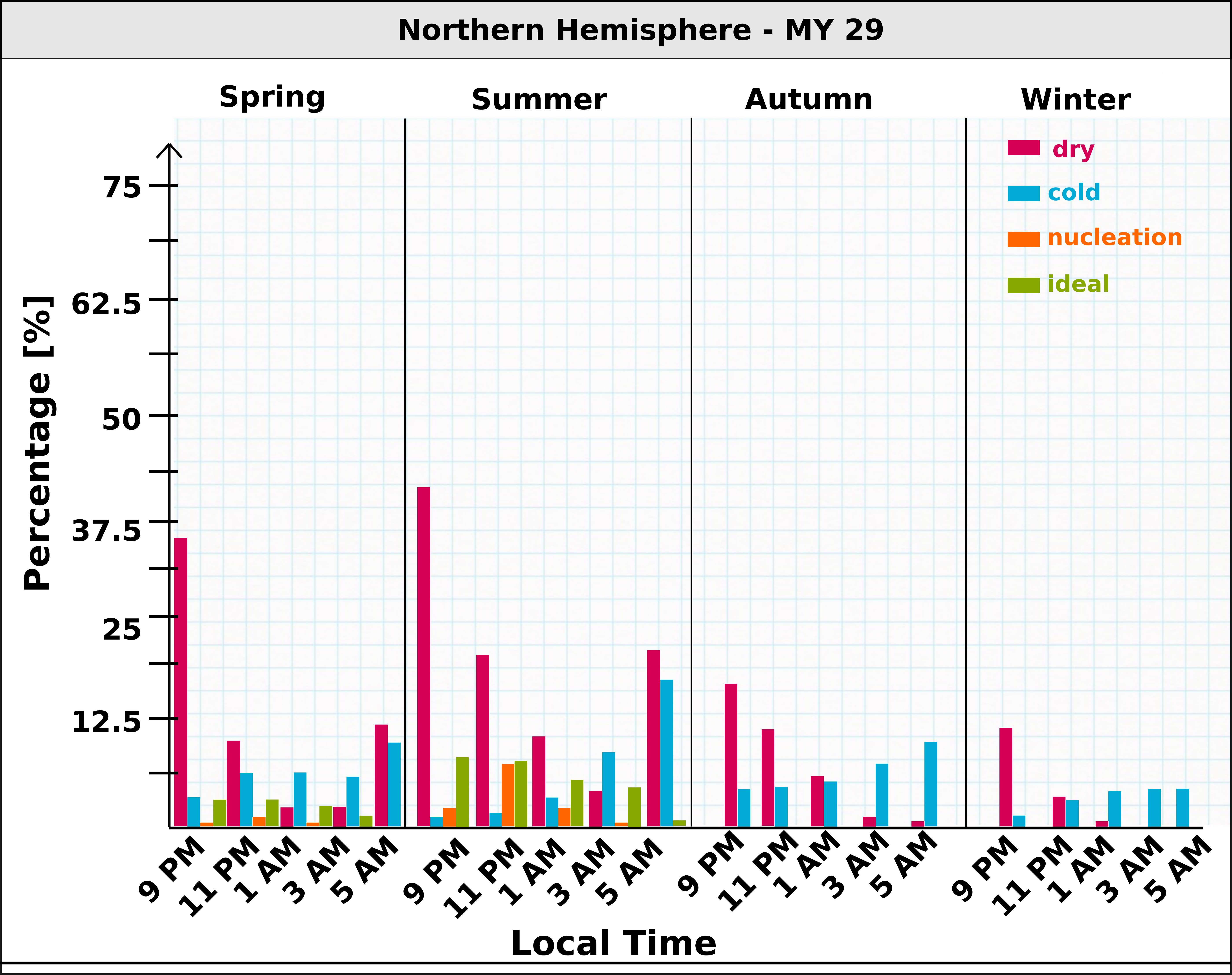}
    \caption{Distribution of different non-ideal and ideal cases of calcium perchlorate deliquescence in the northern hemisphere of Mars, between 0$^{\circ}$ - 80$^{\circ}$ N. Dry: relative humidity with respect to  liquid too low. Cold: surface temperature too low. Nucleation: relative humidity with respect to  ice too high. Ideal: possible chance for deliquescence.}
    \label{fig:nh}
\end{figure}

\begin{figure}
    \centering
    \includegraphics[width=\linewidth]{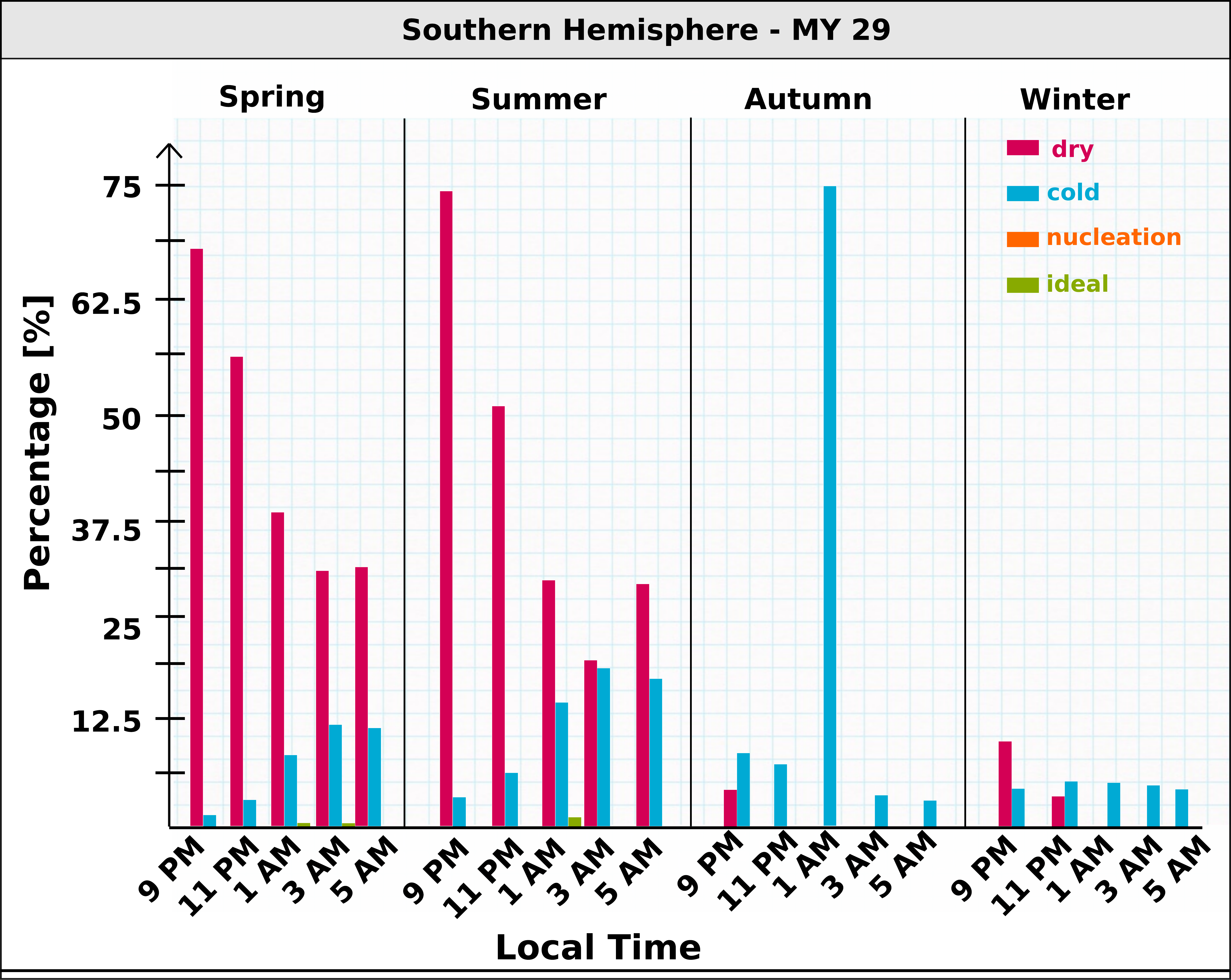}
    \caption{Distribution of different non-ideal and ideal cases of calcium perchlorate deliquescence in the southern hemisphere of Mars, between 0$^{\circ}$ - 80$^{\circ}$ S. Dry: relative humidity with respect to  liquid too low. Cold: surface temperature too low. Nucleation: relative humidity with respect to  ice too high. Ideal: possible chance for deliquescence.}
    \label{fig:sh}
\end{figure}

In this subsection we consider the cases, when all the necessary circumstances for deliquescence were met except for one. Summing these instances up and taking a look at the statistics, we can get a clearer picture of the limiting circumstances according to local time and location (Fig. \ref{fig:nh} and \ref{fig:sh}). We calculated the percentages with respect to every single location between 80$^{\circ}$N - 80$^{\circ}$S and local time (the polar regions were cut due to frequent saturation in the model). An example calculation would be: 35 \% of dry scenarios mean that compared to all instances (43 latitude gridpoints, 65 longitude gridpoints and 669 sols totaling 1 869 855 cases at each local time between 9 PM and 5 AM), 35 \% of them (approximately 654 449) fell into the dry scenario. \\ 

If we examine the Mars year 29 and separate the data by hemispheres and by seasons, the data suggests that relative humidity is the primary limiting factor most of the time. Albeit the temperature is above the necessary eutectic temperature, the relative humidity remains too low (dry scenario). This is a substantially more frequent case, than when either the relative humidity with respect to liquid is high enough with too low temperatures (cold scenario), or the relative humidity with respect to ice is too high, favoring nucleation instead of deliquescence (nucleation scenario). \\

\begin{figure}
    \centering
    \includegraphics[width=\linewidth]{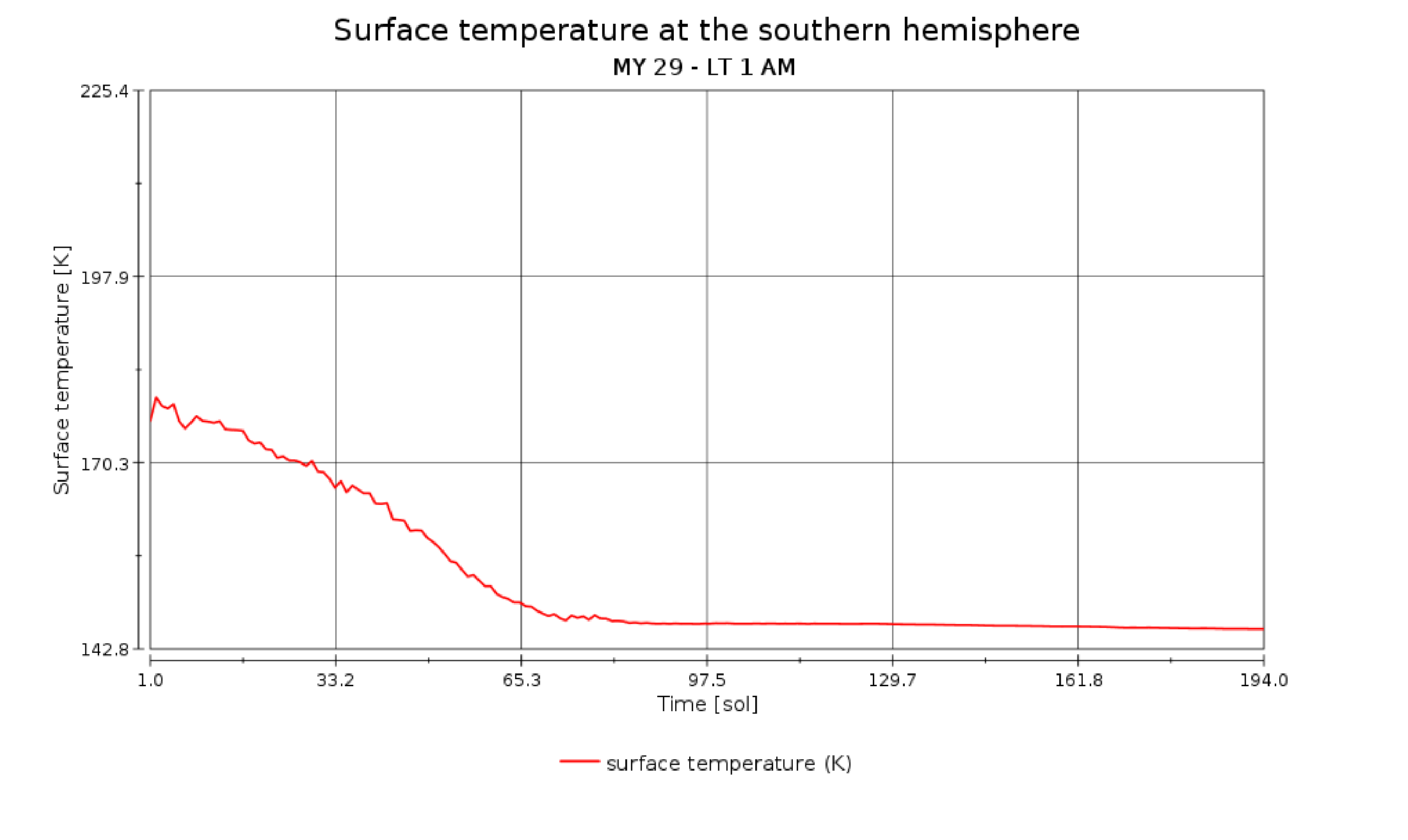}
    \caption{Surface temperature in the southern hemisphere during the local autumn. }
    \label{fig:tssh}
\end{figure}

\begin{figure}
    \centering
    \includegraphics[width=\linewidth]{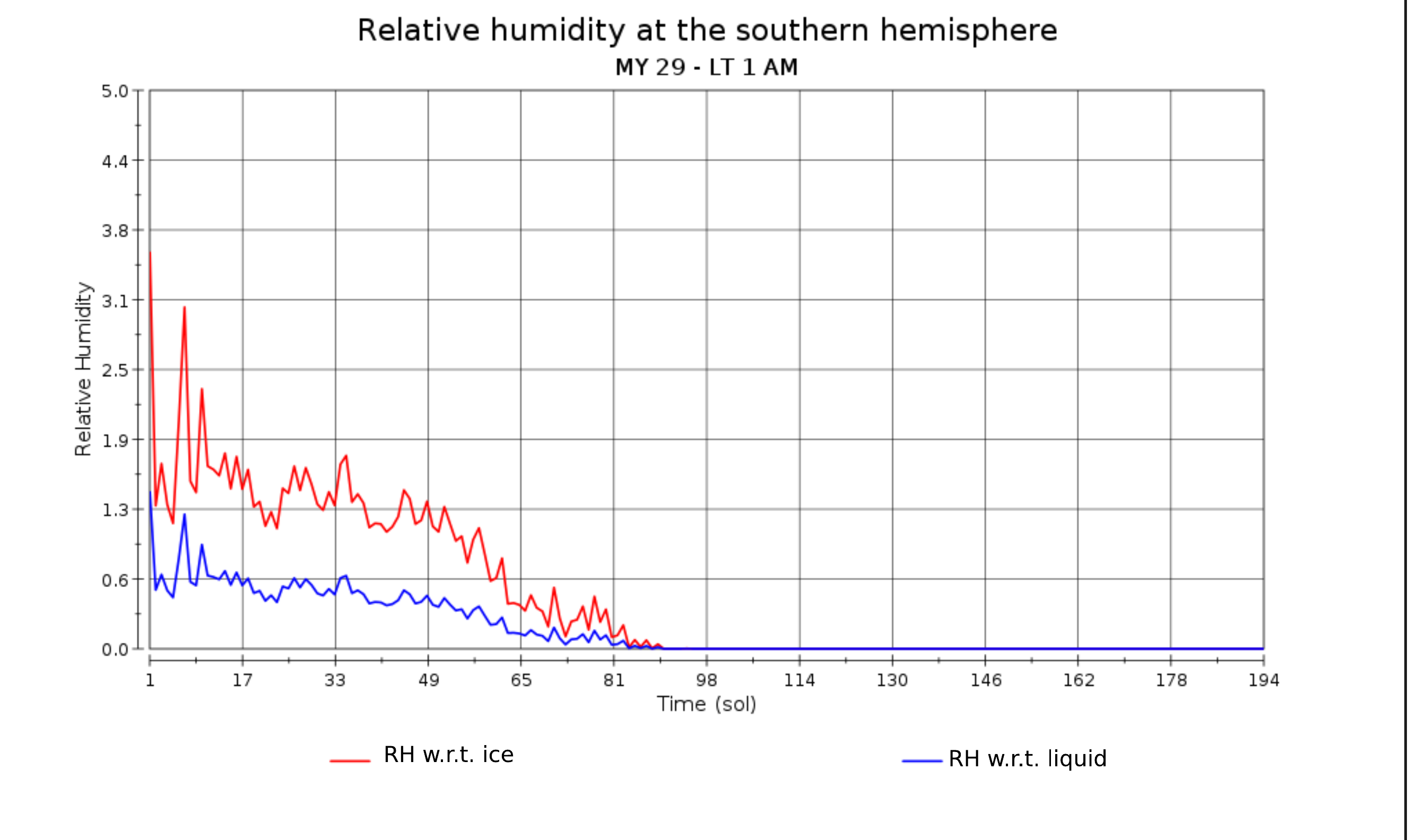}
    \caption{Relative humiditiy with respect to  liquid (blue) and with respect to  ice (red) in the southern hemisphere during the local autumn. }
    \label{fig:rhsh}
\end{figure}

Separating by seasons, the local spring and summer show higher percentages of being too dry with sufficient temperatures. Besides, the local spring and summer in the northern hemisphere display the highest chance for deliquescence (ideal scenario) between 9 PM and 3 AM. There is a similar trend in the southern hemisphere, but with overall lower percentages and mostly between 11 PM and 3 AM. In the southern hemisphere from 1 AM to 5 AM the ratio for too cold scenarios also increase, resulting in a 8-15 \% chance. There is one particularly interesting point, which is 1 AM in the southern hemisphere, during the local autumn. At this time the ratio for too cold scenarios is 74.6 \%, an obvious outlier compared to all the other cold cases. If we take a look at a close-up between sol 1 and sol 194 at the southern hemisphere at 60$^{\circ}$S, 1 AM local time, the relative humidity is elevated between sol 1 and sol 80, the beginning of the local autumn (Fig. \ref{fig:rhsh}). This elevation is visible at all latitudes, 60$^{\circ}$S was selected as an illustration. Comparing the 1 AM data with 11 PM and 3 AM data, in the latter cases the relative humidity is still elevated at the beginning of the year, but not enough to stay above the necessary minimum level for calcium perchlorate. At the same time, the surface temperature remains strictly below the eutectic temperature (Fig. \ref{fig:tssh}), thus we arrive at the 74.6 \% of cold scenarios. \\

Overall the data suggests that relative humidity could be the main limiting factor to deliquescence during local spring and summer in both hemispheres. The ratios, where surface temperature was high enough with relative humidity with respect to ice still low enough is considerably higher, than the other scenarios (too cold or nucleation). The northern hemisphere seems more favourable between 9 PM and 3 AM, while in the southern hemisphere between 1 AM and 3 AM are suggested as more ideal local times. From the hemispherically and seasonally averaged data we can conclude, that the overall chance for ideal periods is still rather low, but usually not zero. Reviewing local ``hot spots for deliquescence'' might be a good follow up to these results. \\

\section{Discussion and conclusions}
\label{sec:discussion}

As the search for extraterrestrial habitability (past or present) and even future manned mission plans (requiring in-situ water production) are gaining popularity, the importance of searching and predicting liquid water formation and the stability of solutions on the surface of Mars is increasing. Deliquescence is a way for liquid water to form, even in present conditions, that shows good potential, but is not yet widely researched in Martian conditions. The investigation of annual, global distribution of potential briny liquids is a substantial objective, that guides the selection of future landing sites with the aim of finding or producing/acquiring liquid water from local sources. \\

We used meteorological modelling to find seasonal and annual characteristics of calcium perchlorate and magnesium perchlorate salt deliquescence on Mars. We followed the criteria for liquefaction detailed in the work of \citet{rivera2018}, and used an equation based on the Goff-Gratch equation to calculate relative humidities (with respect to ice and with respect to liquid) as previously described in \citet{pal2020}. In general, the late night hours, from 9 PM to 3 AM, showed a significant global chance for deliquescence during the local late spring and early summer (L$_s$ 90$$^\circ$$ -- 130$$^\circ$$)  in the northern hemisphere. Acidalia Planitia and Utopia Planitia show the best chance for Ca(ClO$_4$)$_2$ deliquescence in the northern hemisphere, while there is chance for deliquescence scattered across the northern upper latitudes (as can be seen in the animations included in the online version). While there is no significant difference in the daily maximum temperatures between these two regions and the rest of the northern hemisphere below 60$^\circ$, the minimum temperatures are 30-50 K higher. The overall higher amounts of water vapor combined with higher night time temperatures could be the reason why these two larger regions show up in the annual average figures. The lower thermal inertia regions, that showed elevated relative humidity values throughout the year, seem to be less ideal for the deliquescence process \citep{pal2019}. \\

In the southern hemisphere the best season was also local spring (L$_s$ 230$$^\circ$$ -- 300$$^\circ$$) with smaller chance than in the northern, produced by the increased atmospheric water vapor through exposed water ice from below the receding seasonal carbon-dioxide polar cap. At the southern hemisphere Hellas Planitia and Argyre Planitia both show some chance for deliquescence, with 3 AM being the overall best period during the year. They both have higher daily minimum temperatures compared with their vicinity, with no significant variation in relative humidity levels \citep{pal2019}.   \\

Patterns visible in annual zonal average deliquescence probability figures largely follow the annual cycle of water vapor \citep{smith2002}. The northern upper arch starting around sol 67 is in agreement with the seasonal sublimation of the CO$_2$ ice cap, releasing water vapor from below. In both hemispheres the amount of atmospheric water vapor reaches maximum during the local summertime, however, after the decay of this maximum their behaviour differs completely. In the northern hemisphere, between mid-summer and mid-autumn, the area of relatively high water vapor extends increasingly towards the equator - crossing over to the southern hemisphere eventually. This extension towards the equator is visible in the zonal average maps at all local times. In comparison, there is no significant transport from the southern hemisphere to the northern hemisphere. The amplitude of the water vapor maxima differs as well, with approximately 85\% of the total atmospheric water vapor in the northern hemisphere at maximum, and only about 65\% in the southern hemisphere at maximum. Possible deliquescence area is substantially larger in the northern hemisphere, that could be the result of the overall higher amounts of water vapor in the atmosphere. \\

The outer edges of the zonally averaged ideal regions are in strong agreement with the seasonal recession of the polar ice caps (see \citet{james1990, schmidt2009} for the southern, and \citet{james2001, calvin2015} for the northern ice caps). As the ice caps sublimate in warmer seasons, releasing water vapor back into the atmosphere, the relative humidity levels reach the necessary minimum for deliquescence. In the northern hemisphere there is a chance at almost all latitudes around 9 PM and 11 PM, the ideal regions narrowing down to Acidalia Planitia and parts of Utopia Planitia by the early morning hours. However, comparing the two hemispheres, the potentially ideal regions are substantially smaller in the southern hemisphere. The narrow ideal ``curved band-like'' feature in the southern hemisphere is essentially showing the recession and refreezing of the southern polar cap. As the CO$_2$ ice sublimates, the albedo of the surface decreases, thus the daily surface temperatures get higher \citep{wagstaff2008,appere2011}. Due to the difference of the freezing point of H$_2$O and CO$_2$, the H$_2$O is the first to freeze to the surface, followed by the CO$_2$. During the southern spring the CO$_2$ ice starts to sublime, exposing the frozen H$_2$O layer. Above this layer, the atmospheric water vapor might be locally higher, thus creating the potentially ideal circumstances for deliquescence for a short time. Taking the temperature, available water vapor and overall meteorological differences into account, the deliquescence chance variations could be anticipated. \\

If there is a modelling based theoretical possibility for deliquescence, it does not mean it happens in reality. However such modelling based calculations are useful, partly as there are no better methods available for the estimation of deliquescence yet, and partly because these calculations fine tune the temporal and spatial characteristics where instruments on next missions should focus; like the HABIT \citep{torres2018, martintorres2020,nazarious2019} onboard the ExoMars rover mission \citep{vago2017,vago2019}. In this work we presented the meteorological conditions based evaluation of deliquescence, which requires the presence and hydration state of calcium perchlorate on the surface. Depending on the hydration state of calcium perchlorate, the threshold limit of deliquescence might also change. Based on the work of \citet{nuding2014}, higher hydration states deliquesce at a higher relative humidty than the lower hydration states.\\  

Taking the thermal inertia (TI) values into consideration, the ideal regions have overall higher values, in the range of 170 - 250 Jm$^{-2}$K$^{-1}$s$^{-1/2}$ (indicating fine grained particles as the surface layer), with the coarser material bearing (> 400 Jm$^{-2}$K$^{-1}$s$^{-1/2}$) upper regions of Acidalia Planitia and Utopia Planitia showing the best chance during the late night hours. In the southern hemisphere, parts of Argyre Basin and Hellas Basin are covered with slightly higher TI materials, but the contrast with surrounding regions is significantly less, than in the northern hemisphere. As a result the low TI and the fine material causing it, this zone is unfavourable because it leads to too low nighttime temperatures. Considering the band around 60$^\circ$ S, the TI values ``shift'' here from the 170 - 250 Jm$^{-2}$K$^{-1}$s$^{-1/2}$ range to the low 65 - 165 Jm$^{-2}$K$^{-1}$s$^{-1/2}$ range indicating a dusty, very fine layer. This band corresponds most likely to the H$_2$O ``ring'' around the receding seasonal southern polar ice cap. \\


There is limited amount of research regarding deliquescence mapping, but a map showing the distribution of (meta)stable brines was published recently, thanks to the work of \citet{rivera2020}. Comparing our results to their work, the potential regions are a good match both in the case of Ca(ClO$_4$)$_2$ and Mg(ClO$_4$)$_2$. The almost continuous band above 30$^\circ$ N (most visible at 9 PM in our results) is also present in both studies, as well as the southern band around 60$^\circ$ S (most visible at 1 AM in our results). One difference to highlight is Argyre Planitia, which showed a small chance of deliquescence in our results, but is an area, where such brines cannot form in the work of \citet{rivera2020}. Thus, Argyre Planitia would be an interesting place for further research to see, what may cause the differences in results. \\

The driving force of the enigmatic Recurring Slope Lineae (RSL) is another highly debated question in the topic of possible signs of martian liquid water. Perchlorates were detected near some RSL sites \citep{ojha2015}, thus the hypothesis of brine flow produced RSLs \citep{gough2016} gained new momentum. \citet{mitchell2016} tested this theory, and concluded that deliquescence of water from the atmosphere is an unlikely mechanism for RSL formation due to the atmospheric water vapor not reaching high enough quantities. The distribution of chlorides and confirmed RSL sites showed in their work range from approximately 45$^\circ$ S to the equator. If we compare this to our maps we can see, that our results reinforce, that even with confirmed sources of chlorides, the region is not likely to be an ideal place for deliquescence. \citet{munaretto2020} presented the first CaSSIS (Color and Surface Science Imaging System on board ESA's ExoMars Trace Gas Orbiter) observation of RSL at Hale crater during the early morning hours, and their results also support dry granular flow dynamics \citep{dundas2017, vincendon2019, schaefer2019}. Although Argyre Basin shows a chance for deliquescence in our results, Hale crater, located just north, does not appear to be ideal, thus aligning with the conclusion of \citet{munaretto2020}. In contrast, in the recent work of \citet{horne2018} the dry flow mechanisms seemed inadequate to explain the leeved fissures and gutters in Meridinai Planum. These fissures and gutters could provide geomorphological evidence of recent transient processes, for example deliquescence of calcium perchlorate. In our results, the equatorial region shows limited chance for deliquescence, however, Meridiani Planum does seem to show a small chance throughout the year.  \\

The results we presented in this work were calculated using the LMDZ GCM, and as such has certain limitations. The saturation of the northern polar region could be an artifact related to the seasonal evaporation of the polar ice, rather than a definite chance for deliquescence. In our work we assume the water vapor volume mixing ratio to be well mixed between approximately 4 meters altitude and at the surface. This is due to the complexity of near-surface processes, resulting in uncertain water vapor data under 4 meters (the first altitude level in the atmospheric model). This approach gives us reasonable results and is currently the best way to estimate the relative humidity this close to the surface from the GCM. Analysing temporal humidity changes might help to improve this, from the data of future missions.\\

We have calculated for the Mars year 29 in this work, and comparing our results with different years would be a beneficial direction for a follow-up study. Analyzing the differences and similarities between multiple martian years would be a great way to distinguish year specific results. Another compelling question to pursue could be a detailed study of the chance for deliquescence in the Argyre and Hellas Planitia regions. Taking hygroscopic salts with different eutectic temperature - relative humidity pairs into consideration would be another good path to better understand the balance of temperature and relative humidity in global deliquescence probabilities.  \\

Our results suggest that there is a good chance for calcium perchlorate deliquescence at larger regions on Mars, and the possibly ideal times are spread throughout several hours. From the statistics of our calculations, relative humidity might be the main limiting factor of deliquescence. Our deliquescence probability maps could be used as a good starting point in finding interesting locations for local ephemeral brine studies, or as a preliminary guide in planning future, liquid brine related in-situ missions. \\  

\section{Acknowledgement}
\label{sec:ack}

This work was supported by the EXODRILTECH project of ESA and the Excellence of Strategic R\&D centres (GINOP-2.3.2-15-2016-00003) project of NKFIH and the related H2020 fund, the COOP-NN-116927 project of NKFIH and the TD1308 \textit{Origins and evolution of life on Earth and in the Universe} COST actions number 39045 and 39078. The NetCDF files were visualized with the NASA GISS Panoply viewer developed by Dr. Robert B. Schmunk at the NASA Goddard Institute for Space Studies. \\

\appendix

\section{Limiting factors of deliquescence}
\label{appendix-stats}

\begin{table}[htbp]
\begin{tabular}{|l|c|c|r|r|r|r|}
\hline
\multicolumn{1}{|c|}{Local time} & Lat & Sol & \multicolumn{1}{c|}{Cold} & \multicolumn{1}{c|}{Nucleation} & \multicolumn{1}{c|}{Dry} & \multicolumn{1}{c|}{Ideal} \\ \hline
\multicolumn{ 1}{|c|}{09:00 PM} & \multicolumn{ 1}{c|}{0 $^{\circ}$ - 80 $^{\circ}$} & 1 – 194 & 4.332\% & 1.076\% & 35.264\% & 4.065\% \\ \cline{ 3- 7}
\multicolumn{ 1}{|l|}{} & \multicolumn{ 1}{c|}{} & 194 – 372 & 2.648\% & 4.599\% & 42.418\% & 8.821\% \\ \cline{ 3- 7}
\multicolumn{ 1}{|l|}{} & \multicolumn{ 1}{c|}{} & 372 – 514 & 5.454\% & 0.000\% & 15.678\% & 0.011\% \\ \cline{ 3- 7}
\multicolumn{ 1}{|l|}{} & \multicolumn{ 1}{c|}{} & 514 – 669 & 3.926\% & 0.000\% & 11.248\% & 0.004\% \\ \hline
\multicolumn{ 1}{|c|}{09:00 PM} & \multicolumn{ 1}{c|}{-80 $^{\circ}$ - 0 $^{\circ}$} & 1 – 194 & 7.190\% & 0.004\% & 4.492\% & 0.018\% \\ \cline{ 3- 7}
\multicolumn{ 1}{|l|}{} & \multicolumn{ 1}{l|}{} & 194 – 372 & 5.848\% & 0.002\% & 9.186\% & 0.089\% \\ \cline{ 3- 7}
\multicolumn{ 1}{|l|}{} & \multicolumn{ 1}{l|}{} & 372 – 514 & 2.034\% & 0.014\% & 68.404\% & 0.090\% \\ \cline{ 3- 7}
\multicolumn{ 1}{|l|}{} & \multicolumn{ 1}{l|}{} & 514 – 669 & 3.120\% & 0.005\% & 71.592\% & 0.116\% \\ \hline
\multicolumn{ 1}{|c|}{11:00 PM} & \multicolumn{ 1}{c|}{0 $^{\circ}$ - 80 $^{\circ}$} & 1 – 194 & 7.122\% & 3.842\% & 10.630\% & 4.744\% \\ \cline{ 3- 7}
\multicolumn{ 1}{|l|}{} & \multicolumn{ 1}{c|}{} & 194 – 372 & 3.762\% & 8.581\% & 17.741\% & 8.642\% \\ \cline{ 3- 7}
\multicolumn{ 1}{|l|}{} & \multicolumn{ 1}{c|}{} & 372 – 514 & 5.861\% & 0.000\% & 11.077\% & 0.002\% \\ \cline{ 3- 7}
\multicolumn{ 1}{|l|}{} & \multicolumn{ 1}{c|}{} & 514 – 669 & 5.059\% & 0.000\% & 4.669\% & 0.000\% \\ \hline
\multicolumn{ 1}{|c|}{11:00 PM} & \multicolumn{ 1}{c|}{-80 $^{\circ}$ - 0 $^{\circ}$} & 1 – 194 & 6.704\% & 0.001\% & 0.750\% & 0.004\% \\ \cline{ 3- 7}
\multicolumn{ 1}{|l|}{} & \multicolumn{ 1}{l|}{} & 194 – 372 & 6.332\% & 0.000\% & 3.101\% & 0.016\% \\ \cline{ 3- 7}
\multicolumn{ 1}{|l|}{} & \multicolumn{ 1}{l|}{} & 372 – 514 & 4.205\% & 0.140\% & 55.897\% & 0.479\% \\ \cline{ 3- 7}
\multicolumn{ 1}{|l|}{} & \multicolumn{ 1}{l|}{} & 514 – 669 & 6.616\% & 0.033\% & 52.027\% & 0.662\% \\ \hline
\multicolumn{ 1}{|c|}{01:00 AM} & \multicolumn{ 1}{c|}{0 $^{\circ}$ - 80 $^{\circ}$} & 1 – 194 & 7.120\% & 1.966\% & 3.635\% & 3.091\% \\ \cline{ 3- 7}
\multicolumn{ 1}{|l|}{} & \multicolumn{ 1}{c|}{} & 194 – 372 & 5.663\% & 4.694\% & 7.038\% & 6.885\% \\ \cline{ 3- 7}
\multicolumn{ 1}{|l|}{} & \multicolumn{ 1}{c|}{} & 372 – 514 & 6.477\% & 0.000\% & 6.025\% & 0.000\% \\ \cline{ 3- 7}
\multicolumn{ 1}{|l|}{} & \multicolumn{ 1}{c|}{} & 514 – 669 & 6.680\% & 0.000\% & 1.724\% & 0.000\% \\ \hline
\multicolumn{ 1}{|c|}{01:00 AM} & \multicolumn{ 1}{c|}{-80 $^{\circ}$ - 0 $^{\circ}$} & 1 – 194 & 74.661\% & 0.000\% & 0.063\% & 0.000\% \\ \cline{ 3- 7}
\multicolumn{ 1}{|l|}{} & \multicolumn{ 1}{l|}{} & 194 – 372 & 6.601\% & 0.000\% & 0.861\% & 0.005\% \\ \cline{ 3- 7}
\multicolumn{ 1}{|l|}{} & \multicolumn{ 1}{l|}{} & 372 – 514 & 8.046\% & 0.465\% & 39.016\% & 1.662\% \\ \cline{ 3- 7}
\multicolumn{ 1}{|l|}{} & \multicolumn{ 1}{l|}{} & 514 – 669 & 12.765\% & 0.137\% & 30.434\% & 2.237\% \\ \hline
\multicolumn{ 1}{|c|}{03:00 AM} & \multicolumn{ 1}{c|}{0 $^{\circ}$ - 80 $^{\circ}$} & 1 – 194 & 6.419\% & 0.528\% & 3.072\% & 2.419\% \\ \cline{ 3- 7}
\multicolumn{ 1}{|l|}{} & \multicolumn{ 1}{c|}{} & 194 – 372 & 8.776\% & 1.171\% & 4.913\% & 5.698\% \\ \cline{ 3- 7}
\multicolumn{ 1}{|l|}{} & \multicolumn{ 1}{c|}{} & 372 – 514 & 8.720\% & 0.000\% & 2.956\% & 0.000\% \\ \cline{ 3- 7}
\multicolumn{ 1}{|l|}{} & \multicolumn{ 1}{c|}{} & 514 – 669 & 6.795\% & 0.000\% & 0.337\% & 0.000\% \\ \hline
\multicolumn{ 1}{|c|}{03:00 AM} & \multicolumn{ 1}{c|}{-80 $^{\circ}$ - 0 $^{\circ}$} & 1 – 194 & 4.303\% & 0.000\% & 0.001\% & 0.0004\% \\ \cline{ 3- 7}
\multicolumn{ 1}{|l|}{} & \multicolumn{ 1}{l|}{} & 194 – 372 & 5.865\% & 0.000\% & 0.250\% & 0.0004\% \\ \cline{ 3- 7}
\multicolumn{ 1}{|l|}{} & \multicolumn{ 1}{l|}{} & 372 – 514 & 11.445\% & 0.030\% & 29.070\% & 1.242\% \\ \cline{ 3- 7}
\multicolumn{ 1}{|l|}{} & \multicolumn{ 1}{l|}{} & 514 – 669 & 18.917\% & 0.001\% & 20.519\% & 0.850\% \\ \hline
\multicolumn{ 1}{|c|}{05:00 AM} & \multicolumn{ 1}{c|}{0 $^{\circ}$ - 80 $^{\circ}$} & 1 – 194 & 9.896\% & 0.006\% & 11.341\% & 0.526\% \\ \cline{ 3- 7}
\multicolumn{ 1}{|l|}{} & \multicolumn{ 1}{c|}{} & 194 – 372 & 14.222\% & 0.036\% & 17.930\% & 1.898\% \\ \cline{ 3- 7}
\multicolumn{ 1}{|l|}{} & \multicolumn{ 1}{c|}{} & 372 – 514 & 10.278\% & 0.000\% & 1.465\% & 0.000\% \\ \cline{ 3- 7}
\multicolumn{ 1}{|l|}{} & \multicolumn{ 1}{c|}{} & 514 – 669 & 6.579\% & 0.000\% & 0.003\% & 0.000\% \\ \hline
\multicolumn{ 1}{|c|}{05:00 AM} & \multicolumn{ 1}{c|}{-80 $^{\circ}$ - 0 $^{\circ}$} & 1 – 194 & 4.001\% & 0.000\% & 0.000\% & 0.000\% \\ \cline{ 3- 7}
\multicolumn{ 1}{|l|}{} & \multicolumn{ 1}{l|}{} & 194 – 372 & 5.294\% & 0.000\% & 0.039\% & 0.000\% \\ \cline{ 3- 7}
\multicolumn{ 1}{|l|}{} & \multicolumn{ 1}{l|}{} & 372 – 514 & 10.403\% & 0.000\% & 29.802\% & 0.043\% \\ \cline{ 3- 7}
\multicolumn{ 1}{|l|}{} & \multicolumn{ 1}{l|}{} & 514 – 669 & 15.130\% & 0.005\% & 27.884\% & 0.014\% \\ \hline
\end{tabular}
\caption{Statistics of the limiting factor in deliquescence, detailed in Section \ref{subsec:limit}.}
\label{tab:stats}
\end{table}



\bibliographystyle{elsarticle-harv}

\bibliography{DeliqMap_Review.bib}

\end{document}